\documentclass[twocolumn,showpacs,preprintnumbers,amsmath,amssymb]{revtex4}

\usepackage{graphicx}
\usepackage{dcolumn}
\usepackage{bm}

\begin{document}

\preprint{APS/123-QED}

\title{Variation of the speed of light with temperature of the expanding universe}
\author{Cl\'audio Nassif and A. C. Amaro de Faria Jr.\\
 cnassif172@gmail.com, antoniocarlos@ieav.cta.br}

 \altaffiliation{{\bf UFOP}: Retired professor of Universidade Federal de Ouro Preto, residence in Rua
 Rio de Janeiro 1186, 1304, 30160-041, Belo Horizonte-MG, whose name is CPFT (Centro de Pesquisas em 
 F\'isica Te\'orica: {\bf a non-profit} name of fantasy), Brazil.\\
  {\bf IEAv}: Instituto de Estudos Avan\c{C}dos, Rodovia dos Tamoios Km 099, 12220-000, S\~ao Jos\'e dos Campos-SP, Brazil.}

\date{\today}

\begin{abstract}

  From an extended relativistic dynamics for a particle moving in a cosmic background field with temperature $T$, we aim to obtain the
 speed of light with an explicit dependence on the background temperature of the universe. Although finding the speed of light in the early
 universe much larger than its current value, our approach does not violate the postulate of special relativity. Moreover, it is shown that
 the high value of the speed of light in the early universe was drastically decreased before the beginning of the inflationary period. So 
 we are led to conclude that the theory of varying speed of light should be questioned as a possible solution of the horizon problem. 

\end{abstract}

\pacs{03.30.+p, 11.30.Qc}
\maketitle

\section{\label{sec:level1} Introduction}
The advent of varying speed of light (VSL) theories\cite{1}\cite{2}\cite{3}\cite{4}\cite{5}\cite{6} seems to shake the
foundations of special relativity theory because the speed of light $c$ in vacuum is no longer constant. However, we must take 
care to investigate the veracity of such proposals. For this, we will take into account an extended relativistic dynamics due to the 
presence of an isotropic background field with temperature $T$, where we consider that the energy scale at which a particle is subjected 
has a nonlocal origin, representing the background thermal energy of the whole universe; i.e., the particle should be coupled to the
background field with temperature $T$, 
which leads to a correction on its energy. This result will allow us to obtain the speed of light with an explicit dependence on the
temperature of the universe. So we will be able to preserve the postulate of constancy of the speed of light and extend it just for the 
implementation of the temperature of the expanding universe. In this sense, we have a function $c(T)$ and so we will find an enormous
value for the speed of light in the early universe when its temperature was extremely high. In addition, we will also conclude that the 
high speed of light was rapidly damped to a value much closer to its current value even before the beginning of the cosmic inflation. This
result will lead us to question VSL theory as an alternative explanation for the horizon problem.

 \section{\label{sec:level1} Energy equation of a particle in a thermal background field}
According to the relativistic dynamics, the relativistic mass of a particle is $m=\gamma m_0$, where $\gamma=1/\sqrt{1-v^2/c^2}$ and
$m_0$ is its rest mass. On the other hand, according to Newton second law applied to its relativistic momentum, we find 
$F=dP/dt=d(\gamma m_0 v)/dt=(m_0\gamma^3)dv/dt=m_0(1-v^2/c^2)^{-3/2}dv/dt$, where $m_0\gamma^3$ represents an inertial mass 
($m_i$) that is larger than the relativistic mass $m(=\gamma m_0)$; i.e., we have $m_i>m$. 

The mysterious discrepancy between the relativistic mass $m$($m_r$) and the inertial mass $m_i$ from Newton second law is a controversial
issue\cite{7}\cite{8}\cite{9}\cite{10}\cite{11}\cite{12}\cite{13}. Actually the Newtonian notion about inertia as the resistance to 
acceleration ($m_i$) is not compatible with the relativistic dynamics ($m_r$) in the sense that we generally cannot consider 
$\vec F= m_r\vec a$. An interesting explanation for such a discrepancy is to take into consideration the influence of an isotropic
background field\cite{14} that couples to the particle, by dressing its relativistic mass ($m_r$) in order to generate an effective
(dressed) mass $m^*(=m_{effective})$ working like the inertial mass $m_i(>m_r)$ in accordance with the Newtonian concept of inertia, where
we find $m^*=m_i=\gamma^2 m_r=\gamma^2 m$. In this sense, it is natural to conclude that $m^*$ has a nonlocal origin; i.e., it comes from
a kind of interaction with a background field connected to a universal frame\cite{14}, which is within the context of the ideas of 
Sciama\cite{15}, Schr\"{o}dinger\cite{16} and Mach\cite{17}.

If we define the new factor $\gamma^2=\Gamma$, we write 

\begin{equation}
 m^*=\Gamma m,
\end{equation}
where $\Gamma$ provides a nonlocal dynamic effect due to the influence of a universal background field over the particle moving with
speed $v$ with respect to such a universal frame\cite{16}. According to this reasoning, the particle is not completely free, since its
relativistic energy is now modified by the presence of the whole universe, namely:  

\begin{equation}
E^*= m^*c^2=\Gamma m c^2
\end{equation}

As the modified energy $E^*$ can be thought as being the energy $E$ of the free particle plus an increment $\delta E$
of nonlocal origin, i.e., $E^*=\Gamma E=E+\delta E$, let us now consider that $\delta E$ comes from the thermal background field
of the whole expanding universe instead of simply a dynamic effect of a particle moving with speed $v$ in the background field,
in spite of the fact that there should be an equivalence between the dynamical and thermal approaches for obtaining the modified energy.
To show this, we make the following consideration inside the factor $\Gamma$, namely:

\begin{equation}
\Gamma(v)=\left(1-\frac{v^2}{c^2}\right)^{-1}\equiv\Gamma(T)=\left(1-\frac{\frac{m_Pv^2}{K_B}}{\frac{m_Pc^2}{K_B}}\right)^{-1}, 
\end{equation}
from where we find $\Gamma(T)=(1-T/T_P)^{-1}$, $T$ being the background temperature. $T_P(=m_Pc^2/K_B\sim 10^{32}$K) is the Planck
temperature in the early universe with Planck radius $R_P\sim 10^{-35}$m. $E_P(=m_Pc^2\sim 10^{19}$GeV) is
the Planck energy and $m_P(\sim 10^{-4}$g) is the Planck mass. From the thermal approach, if $T\rightarrow T_P$, $\Gamma(T)$ diverges. 

 Now we simply rewrite (2) as follows: 

   \begin{equation}
 E=\Gamma(T)mc^2=\frac{\gamma m_0c^2}{1-\frac{T}{T_P}}
   \end{equation}

  As the factor $\Gamma(T)$ has a nonlocal origin and is related to the background temperature of the universe, let us admit that this 
factor acts globally on the speed of light $c$, while the well-known factor $\gamma$ acts locally on the relativistic mass of the
particle. In view of this, we should redefine the equation (4) in the following way:

\begin{equation}
 E=[\gamma^{\prime} m_0][\Gamma(T)c^2]=\gamma^{\prime}m_0c^{\prime 2}=mc(T)^2=mc^{\prime 2},
\end{equation}
where now we have $m=\gamma^{\prime}m_0$, so that 

\begin{equation}
 \gamma^{\prime}=\frac{1}{\sqrt{1-\frac{v^2}{c^{\prime 2}}}}
\end{equation}

  And from Eq.(5) we extract

\begin{equation}
 c^{\prime}=c(T)=\frac{c}{\sqrt{1-\frac{T}{T_P}}}, 
\end{equation}
 where $c(T)=\sqrt{\Gamma(T)}c=\gamma_{T} c$, with $\gamma_{T}=1/\sqrt{1-T/T_P}$. 

From Eq.(7) we find that the speed of light was infinite in the initial universe when $T=T_P$. As the universe was expanding 
and getting colder, the speed of light had been decreased to achieve $c(T)\approx c$ for $T<<T_P$. Currently we have $c(T_0)=c$,
with $T_0\approx 2.73$K.

\begin{figure}
\includegraphics[scale=0.45]{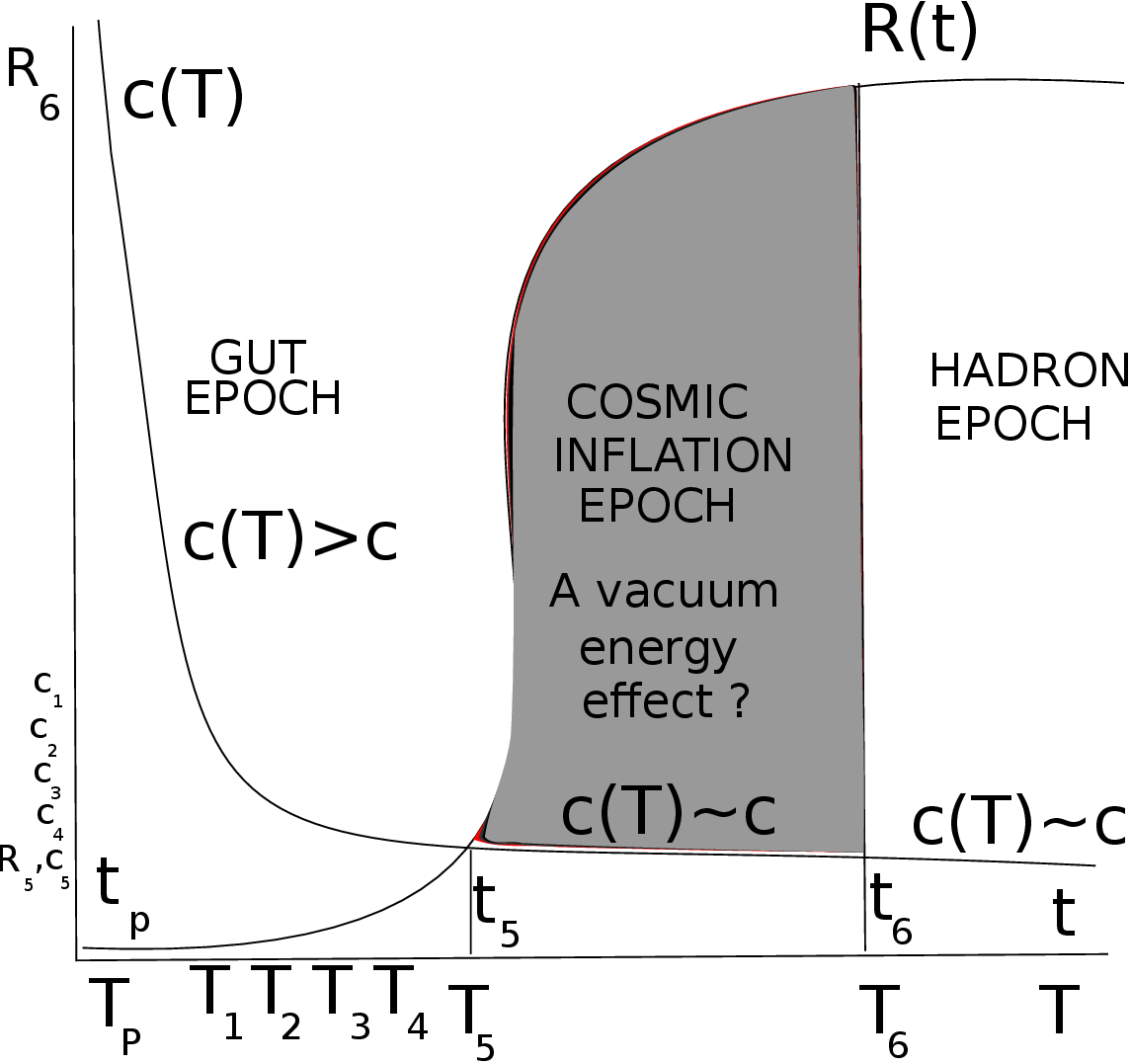}
\caption{This figure shows two graphics, namely $R(t)$, which is the size (radius) of the universe as a function of time, and $c(T)$,  
representing the speed of light with dependence on the temperature of the universe according to Eq.(7). At the beginning of
the universe when it was a singularity with a minimum radius of the order of the Planck radius, i.e., $R_P\sim 10^{-35}$m, 
having the Planck energy scale $E_p\sim 10^{19}$GeV which corresponds to the Planck temperature $T_P\sim 10^{32}$K and the Planck time
$t_P\sim 10^{-43}$s, the speed of light $c^{\prime}$ was infinite since there was no spacetime [see Eq.(7) for $c(T)$].
But immediately after, when $T_1=10^{31}$K, the speed of light had already assumed a value close to the current value as shown by
Eq.(7) for $c(T)$, and therefore a cone of light (a spacetime) had been formed; i.e., with $c=2.99792458\times 10^{8}$m/s for the
present time, then, according to the function $c(T)$, we find $c_1=c(T_1)=3.16008998\times 10^8$ m/s (see the figure). Subsequently,
for $T_2 =10^{30}$K, we find $c_2=c(T_2)=3.01302757\times 10^8$ m/s . For $T_3=10^{29}K\Rightarrow c_3=c(T_3)=2.99942467\times 10^8$m/s.
And for $T_4=10^{28}K\Rightarrow c_4= c(T_4)=2.99807447\times 10^8$m/s. Finally, for $T_5=10^{27}K\Rightarrow c_5=c(T_5)=
2.99793955\times 10^8$m/s. From this temperature $T_5=10^{27}$K, when $t=t_5\sim 10^{-35}$s, corresponding to the energy scale 
of the Grand Unified Theory (GUT) with $10^{14}$GeV, the universe inflated very quickly, starting with a radius 
$R_5\sim 10^{-25}$m and reaching $R_6\sim 10^{25}$m at the time $t_6\sim 10^{-32}$s; i.e., the size of the universe increased rapidly
50 orders of magnitude. Since the speed of light $c_5\approx c$, VSL should be put in doubt. Hence, perhaps the vacuum energy had played
a fundamental role in that epoch.}
\end{figure}

The change in the speed of light is $\delta c=c^{\prime}-c$, namely:

\begin{equation}
 \delta c=(\gamma_{T}-1)c=\left(\frac{1}{\sqrt{1-\frac{T}{T_P}}}-1\right)c,
\end{equation}
where, for $T<<T_P$, we have $\delta c\approx 0$. 

 We should note that the variation of the speed of light with temperature does not invalidate the postulate of constancy of 
the speed of light in special relativity since $c^{\prime}$ for a given temperature remains invariant only with respect
to the motion of massive particles, but not with respect to the age and temperature of the universe.
In other words, we say that, although the speed of light has decreased rapidly during the initial expansion of 
the universe and thereafter with a smooth variation as shown in Fig.1, its value for a given temperature ($c(T)$) 
is still a maximum limit of speed that is invariant only with respect to the motion of all subluminal particles.

 The modified spatial momentum of a particle moving in the presence of a cosmic background field with temperature $T$ is the following: 

\begin{equation}
 P_T=\gamma^{\prime}\gamma_{T} m_0v=\frac{\gamma^{\prime}m_0v}{\sqrt{1-\frac{T}{T_P}}}
\end{equation}

Before obtaining the modified energy-momentum relation, we first introduce the modified 4-velocity, namely, 

\begin{equation}
  U^{\prime\mu}=\left[\frac{\gamma^{\prime}c}{\sqrt{1-\frac{T}{T_P}}}~ , ~
\frac{\gamma^{\prime}v_{\alpha}}{\sqrt{1-\frac{T}{T_P}}}\right],
\end{equation}
where $\mu=0,1,2,3$ and $\alpha=1,2,3$. If $T\rightarrow T_P$, the 4-velocity diverges. 

The modified 4-momentum is $P^{\prime\mu}=m_0U^{\prime\mu}$. So, from (10) we find

\begin{equation}
  P^{\prime\mu}=\left[\frac{\gamma^{\prime}m_0c}{\sqrt{1-\frac{T}{T_P}}}~ , ~
\frac{\gamma^{\prime}m_0v_{\alpha}}{\sqrt{1-\frac{T}{T_P}}}\right],
\end{equation}
where $E=c^{\prime}P^{\prime 0}=m_0\gamma^{\prime}c^{\prime 2}=mc^2/(1-T/T_P)$. The spatial components of $U^{\prime\mu}$ represent the 
spatial momentum given in Eq.(9).

From (11), performing the quantity $P^{\prime\mu}P^{\prime}_{\mu}$, we obtain the modified energy-momentum relation as follows: 

\begin{equation}
 P^{\prime\mu}P^{\prime}_{\mu}=\frac{E^2}{c^{\prime 2}}-P_T^2 = m_0^2 c^{\prime 2},
\end{equation}
from which we get

\begin{equation}
 E^2=\frac{m^2c^4}{\left(1-\frac{T}{T_P}\right)^2}=c^{\prime 2}P_T^2 + m_0^2 c^{\prime 4},
\end{equation}
where $c^{\prime}=c/\sqrt{1-T/T_P}$ [Eq.(7)].

It is curious to notice that the Magueijo-Smolin doubly special relativity equation ($mc^2/1-E/E_P$)\cite{18} reproduces Eq.(4)
when we just replace $E$ by $K_BT$ and $E_P$ by $K_BT_P$ in the denominator of their equation.  

\section{\label{sec:level1} Conclusions}
Figure $1$ reveals to us that, in the very brief time interval between $t_P(\sim 10^{-43}s)$ and $t_5(\sim 10^{-35}s)$, the speed of light
has collapsed from infinite to a finite value $c_5$ very close to its current value $c$ even before the beginning of the period of
inflation where VSL theory attributes a very large speed of light around 60 times its current value. Thus, based on our approach 
by taking into account the background thermal energy, the speed of light can not be used to explain the horizon problem, which 
leads us to think that there was a very rapid expansion of spacetime originating from a vacuum energy that is still poorly understood
\cite{19}. The light has nothing to do with such an effect, since it came into existence first together with the initial singularity. 
All the subsequent cosmic evolution including the very rapid growth of spacetime (inflation) came after the light, whose speed was already
practically established before the cosmic expansion.

{\noindent\bf Acknowledgedments}

 I am grateful to Prof. Fernando A. Silva for very interesting discussions.

\end{document}